\documentclass[superscriptaddress,twocolumn,showpacs,amsmath,amssymb,floats,prl]{revtex4-1}

\usepackage{graphicx}% Include figure filesf

\usepackage{dcolumn}% Align table columns on decimal point

\usepackage{bm}% bold math

\usepackage{revsymb}

\usepackage[usenames]{color}

\usepackage{subfigure}

\usepackage{natbib}

\usepackage{float}

\begin{document}

\newcommand{\brm}[1]{\bm{{\rm #1}}}
\newcommand{\tens}[1]{\underline{\underline{#1}}}
\newcommand{\mm}{\overset{\leftrightarrow}{m}}
\newcommand{\xv}{\bm{{\rm x}}}
\newcommand{\uv}{\bm{{\rm u}}}
\newcommand{\vv}{\bm{{\rm v}}}
\newcommand{\nv}{\bm{{\rm n}}}
\newcommand{\Nv}{\bm{{\rm N}}}
\newcommand{\ev}{\bm{{\rm e}}}
\newcommand{\dv}{\bm{{\rm d}}}
\newcommand{\bv}{\bm{{\rm b}}}
\newcommand{\lv}{{\bm{l}}}
\newcommand{\rv}{\bm{{\rm r}}}
\newcommand{\Rv}{\bm{{\rm R}}}
\newcommand{\piv}{\bm{{\rm \pi}}}
\newcommand{\Av}{\bm{{\rm A}}}
\newcommand{\id}{\tens{\mathbb I}}
\newcommand{\bhv}{\hat{\bv}}
\newcommand{\bh}{\hat{b}}
\def\ten#1{\underline{\underline{{#1}}}}
\newcommand{\Ft}{{\tilde F}}
\newcommand{\Ftv}{\tilde{\mathbf{F}}}
\newcommand{\sigmat}{{\tilde \sigma}}
\newcommand{\sigmab}{{\overline \sigma}}
\newcommand{\ellv}{\mathbf{\ell}}
\newcommand{\qv}{\bm{{\rm q}}}
\newcommand{\pv}{\bm{{\rm p}}}
\newcommand{\tD}{\underline{D}}
\newcommand{\Tchange}[1]{{\color{red}{#1}}}
\newcommand{\Fa}{{\cal F}}
\newcommand{\Uv}{\bm{{\rm U}}}
\newcommand{\Xv}{\bm{{\rm X}}}
\newcommand{\Bv}{\bm{{\rm B}}}
\newcommand{\Dt}{{\tensor {\cal D}}}
\newcommand{\cv}{{\mathbf c}}
\newcommand{\pt}{\tilde{p}}
\newcommand{\ant}[1]{{\color{blue}{#1}}}
\newcommand{\fbz}{1$^{st}$ BZ}
\newcommand{\zvh}{{\bm{{\rm \hat{z}}}}}
\newcommand{\bra}[1]{\langle{{#1}}|}
\newcommand{\ket}[1]{|{{#1}}\rangle}

\title{Organization of strongly interacting directed polymer liquids \\ in the presence of stringent constraints}
\author{Anton Souslov}
\affiliation{School of Physics,
Georgia Institute of Technology, Atlanta, GA 30332, USA }
\author{D. Zeb Rocklin}
\affiliation{Department of Physics,
University of Illinois at Urbana-Champaign, 1110 West Green Street, Urbana, IL 61801, USA}
\author{Paul M. Goldbart}
\affiliation{School of Physics,
Georgia Institute of Technology, Atlanta, GA 30332, USA }

\date{\today}

\begin{abstract}
The impact of impenetrable obstacles on the energetics and equilibrium structure of strongly repulsive directed polymers is investigated. As a result of the strong interactions, regions of severe polymer depletion and excess are found in the vicinity of the obstacle, and the associated free-energy cost is found to scale quadratically with the average polymer density. The polymer-polymer interactions are accounted for via a sequence of transformations: from the 3D line liquid to a 2D fluid of Bose particles to a 2D composite fermion fluid and, finally, to a 2D one-component plasma. The results presented here are applicable to a range of systems consisting of noncrossing directed lines. 
\end{abstract}

%
%Descriptions:

%36.20.Ey 
%30.	ATOMIC AND MOLECULAR PHYSICS
%36.	Exotic atoms and molecules; macromolecules; clusters 
%36.20.-r	Macromolecules and polymer molecules
%36.20.Ey	Conformation (statistics and dynamics)

%11.15.Yc
%10.  THE PHYSICS OF ELEMENTARY PARTICLES AND FIELDS
%11.	General theory of fields and particles
%11.15.-q	 Gauge field theories
%11.15.Yc	Chern-Simons gauge theory

%52.27.Aj
%50. PHYSICS OF GASES, PLASMAS, AND ELECTRIC DISCHARGES
%52. Physics of plasmas and electric discharges
%52.27.-h	Basic studies of specific kinds of plasmas
%52.27.Aj	Single-component, electron-positive-ion plasmas

%52.40.Hf
%50. PHYSICS OF GASES, PLASMAS, AND ELECTRIC DISCHARGES
%52. Physics of plasmas and electric discharges
%52.40.-w Plasma interactions (nonlaser)
%52.40.Hf Plasma-material interactions; boundary layer effects

%05.30.Jp
%00.	GENERAL
%05.  Statistical physics, thermodynamics, and nonlinear dynamical systems
%05.30.-d  	Quantum statistical mechanics
%05.30.Jp  Boson systems (for static and dynamic properties of Bose-Einstein condensates, see 03.75.Hh and 03.75.Kk; see also 67.10.Ba Boson degeneracy in quantum fluids)

%71.10.Pm
%70.	CONDENSED MATTER: ELECTRONIC STRUCTURE, ELECTRICAL, MAGNETIC, AND OPTICAL PROPERTIES
%71.	Electronic structure of bulk materials
%71.10.-w	Theories and models of many-electron systems
%71.10.Pm	Fermions in reduced dimensions (anyons, composite fermions, Luttinger liquid, etc.)

\maketitle

The interaction between dense polymer systems and colloidal particles presents an important and experimentally relevant suite of questions to soft matter science.
Even the issue of a single polymer avoiding a specified obstacle is technically challenging~\cite{Maghrebi2011}. 
Here, our focus is on settings in which interactions \emph{between} the polymers dominates the behavior of the system in the presence of an obstacle.
Prior work in this area includes the study of the polymer free energy and density profile inside a polymer brush due to the inclusion of a colloidal particle~\cite{Kim, Milchev}.

In this Letter, we analytically address the directed polymer liquid --- a strongly interacting system of many thin thermally fluctuating polymers under tension,
which we represent in terms of directed lines in three dimensions.
As with many investigations of ($d$+1)-dimensional systems of directed lines,
our approach hinges on an analogy between the statistical mechanics of these systems
and the quantum mechanics of $d$-dimensional systems of nonrelativistic bosonic particles~\cite{Feynman}.
This analogy has proven useful in other work on directed polymers, either under tension~\cite{DeGennes1968} or within a nematic solvent~\cite{Kamien1992}, as well as work on fluctuating line defects in symmetry-broken phases, such as vortex lines in type-II superconductors~\cite{Nelson1989}.

We consider a system of directed polymers that interact with one another via a strong short-range repulsion, which we treat as a restriction on the possible configurations
of the directed line liquid that prohibits the intersection of any two lines. Central to the present work is the non-perturbative treatment of this nonintersection restriction, which we accomplish
via the well-known Chern-Simons transmutation of quantum statistics, from Bose to Fermi.
Using the quantum many-body physics representation, we address two fundamental questions regarding this system, viz., ``What are the free-energy cost and equilibrium polymer density induced by
{\parfillskip0pt\relax\par}
\begin{figure}[H]
\includegraphics[angle=0]{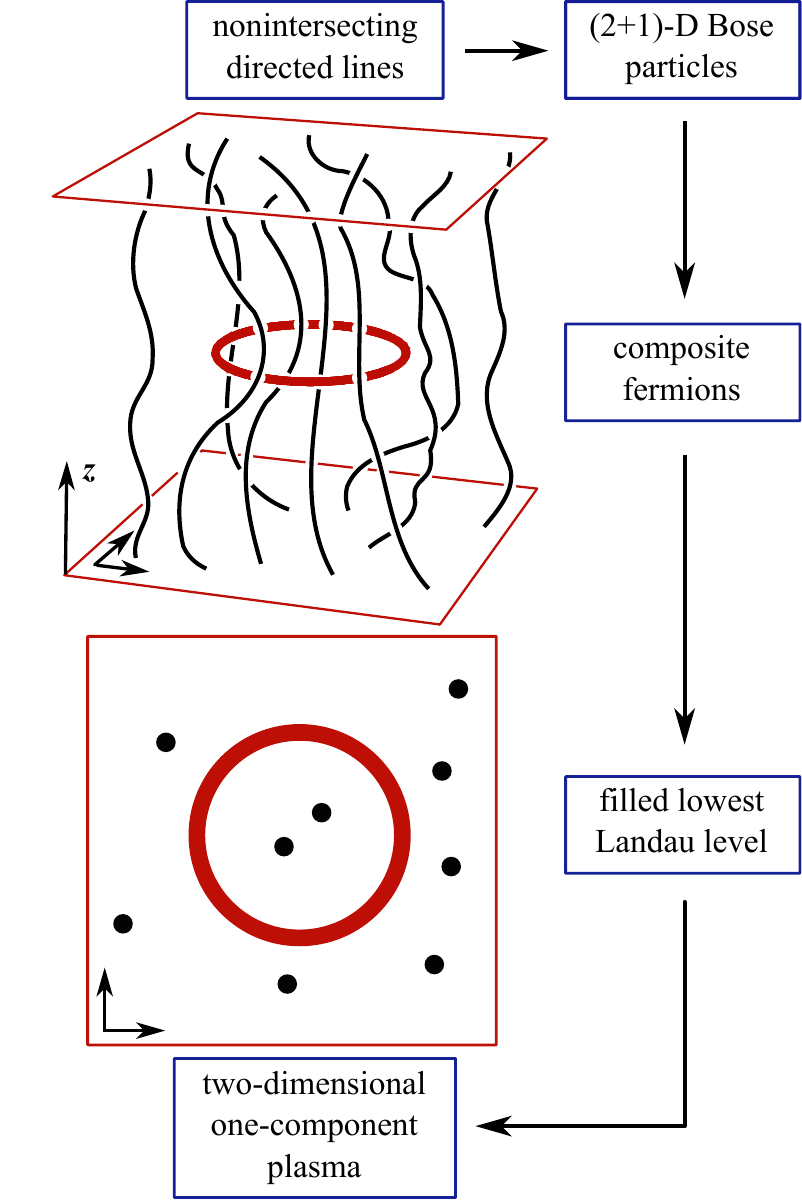}
  \caption{(\emph{Color online}) Sequence of transformations of a liquid of directed polymers subject to a spatially extended constraint, shown as a ring (thick line), threaded by a fixed number of polymers. The sequence is used to obtain the free energy and polymer density profile of the constrained system. Ultimately, the directed polymer liquid is transformed to a two-dimensional one-component plasma, whose the Coulomb repulsion results from integrating out the long noncrossing chains.}
  \label{fig:map}
\end{figure}
\noindent  
 an impenetrable inclusion, i.e., a region in which the polymer density is forced to be zero?'' 
We note that the free-energy cost of the inclusion corresponds to the probability that the system without the inclusion
spontaneously satisfies the condition imposed by an inclusion.
For simplicity, we study thin, flat inclusions within a planar slice perpendicular to 
the polymer direction $\zvh$, and we determine the equilibrium polymer density within that slice. More generally, we determine the free energy and equilibrium density in the presence of spatially extended constraints, such as a ring positioned in the slice and threaded by a fixed number of polymers, as shown in Fig.~\ref{fig:map}. The inclusion is precisely the case for which no polymers thread the ring. We emphasize that the dominant free-energy cost of such constraints is the result of the many-body effect of the crowding of the polymers, an effect absent for free directed lines.

The transformation between directed lines and bosons enables one to harness powerful techniques from quantum many-body physics for application to directed line liquids. 
The first such application was due to de~Gennes, who invoked it to calculate the structure factor for a (1+1)-dimensional hard-core directed line liquid~\cite{DeGennes1968}. A recent elaboration on these results by Rocklin \textit{et al.}~yielded the free energy and equilibrium polymer density in the presence of pins and other impenetrable constraints for the same system~\cite{Rocklin2012}. In essence, de~Gennes's strategy separates into two major steps. First, the classical (1+1)-dimensional noncrossing directed lines are mapped into quantum hard-core point bosons moving in one dimension~\cite{Feynman}. Second, these hard-core bosons are mapped, exactly, to free fermions~\cite{Girardeau1960}. 
The virtue of this sequence of transformations,
which we are here seeking to emulate for (2+1)-dimensional systems of noncrossing 
directed lines, is that it reduces the strongly interacting directed line liquid to an (exactly solvable) system of free fermions. 
We accomplish a similar reduction by mapping our (2+1)-dimensional directed line liquid to a two-dimensional fluid of Bose particles, and then make the Chern-Simons transmutation from Bose to Fermi statistics~\cite{Polyakov1988}.

We address a model of the line liquid that consists of $N$ paths that are directed along the $z$-axis from $z = 0$ to $z = L$ via a line tension $\tau$,
which penalizes deflections of the paths.
We assume no path switchbacks, in which case the (Cartesian) $(x,y)$-coordinates of the $N$ paths $\Rv(z) \equiv  \{\rv_n(z)\}_{n  = 1}^N$ are single-valued functions of $z$.
We study interactions that are short ranged and so
strong that they prohibit configurations in which any pair $(n, n^\prime)$ of lines intersect, and thus $\rv_n(z) \ne \rv_{n^\prime}(z)$ for any value of $z$.
The statistical weight for noncrossing configurations is then
proportional to $\exp\left(-{\cal G}/T\right)$, where 
\begin{equation}
{\cal G}[\Rv(\cdot)] \equiv \frac{\tau}{2} \sum_{n = 1}^N \int_{0}^L \! dz \left| \frac{d \rv_n}{d z} \right|^2,
\end{equation}
and the temperature $T$ is measured in units of the Boltzmann constant.
Correspondingly, the partition function $Z$ is given by the functional integral $Z = \int [D \Rv(\cdot)] e^{- {\cal G}/ T}$, taken over all noncrossing paths.%~\cite{f1}.

To implement the mapping to quantum many-body physics we note that
$Z = \bra{\Psi^F} e^{- \beta L {\cal H}}\ket{\Psi^I}$; see, e.g., Ref.~\cite{Feynman}.
The quantum Hamiltonian, 
\begin{equation}
\label{eq:H}
{\cal H} = \frac{1}{2 m} \sum_{n=1}^N |\pv_n|^2,
\end{equation}
describes two-dimensional Bose particles of mass $m$ under the correspondence
\begin{equation}
\label{eq:corr}
(\tau, T) \longleftrightarrow (m, \hbar).
\end{equation}
In addition, the quantum amplitudes $\langle \Rv \ket{\Psi^{I/F}}$ reflect the initial and final \emph{a priori} classical end-point distributions.
The quantum states $\ket{\Rv}$ are symmetrized products of $N$
single-particle position eigenstates $\ket{\rv}$,
appropriate for indistinguishable polymers.
Observe that the limit $L \to \infty$ is also the low-temperature limit $\beta \to \infty$,
for which the quantum ground state dominates $Z$.

Our focus is on the increase of the free energy $\Delta {\cal F}$ ($ \equiv {\cal F}_c - {\cal F}_u$) from ${\cal F}_u$ (its value in the \emph{unconstrained} directed line liquid) to ${\cal F}_c$ (its value in the 
liquid with a spatially extended \emph{constraint}). We compute $\Delta {\cal F}$ using the ratio 
between the unconstrained partition function $Z$ and 
the partition function $Z_c$, which includes the effects of constraint.
If, as in the case we consider, the constraint requires that exactly $Q$ directed lines pass through some two-dimensional region ${\cal D}$ perpendicular to the director $\zvh$ and
far from the ends $z = 0$, $L$, then, in the limit $L \to \infty$, $\Delta {\cal F}$ may be computed using ground-state dominance as:
\begin{equation}
\label{eq:ds}
\Delta {\cal F} = - T \ln \frac{ Z_c }{ Z } = - T \ln \int_{\cal C} d \Rv \,\, |\Psi_b(\Rv)|^2.
\end{equation}
Here, ${\cal C}$ indicates the constraint on integration that exactly $Q$ of the $N$ coordinates $\{\rv_n\}$ lie within ${\cal D}$.
The ground-state wavefunction in Eq.~(\ref{eq:ds}) $\Psi_b(\Rv)$ ($\equiv \bra{\Rv}\mathrm{GS} \rangle$) is the lowest-energy (particle-exchange symmetric) solution of the energy eigenproblem ${\cal H} \Psi_b = E \Psi_b$. Thus, we are confronted with the task of finding an expression for $|\Psi_b|^2$.

The restrictions $\rv_n(z) \ne \rv_{n^\prime}(z)$ on the path integrals for the line-liquid partition functions $Z$ and $Z_c$
demand a nonperturbative treatment. 
Following the essence of de~Gennes's~\cite{DeGennes1968} strategy,
we transmute the statistics of the quantum fluid from Bose to Fermi~\cite{Polyakov1988}, thus arriving at a description in terms of a many-fermion wavefunction $\Psi_f$ that necessarily has nodes for any two coincident particles, i.e., $\Psi_f(\rv_1, \ldots, \rv, \ldots, \rv, \ldots, \rv_N) = 0$. 
This transmutation is accomplished via the well-known~\cite{Arovas1985} singular gauge transformation of Chern-Simons theory: 
$\Psi_b \to \Psi_f \equiv \Psi_b \exp({i \phi \sum_{n^\prime < n} \theta_{n,n^\prime}})$, where $\theta_{n,n^\prime} \equiv \tan^{-1}\left[(y_n - y_{n^\prime})/(x_n - x_{n^\prime})\right]$ is the polar angle between particles $n$ and $n^\prime$. 
Then, under the exchange of particles $n$ and $n^\prime$, $\theta_{n,n^\prime} \to \theta_{n,n^\prime} \pm \pi$, and, provided $\phi$ is an odd integer, $\Psi_f$ is antisymmetric, i.e., fermionic.
As $\Psi_b$ is the ground state of ${\cal H}$, Eq.~(\ref{eq:H}), $\Psi_f$ obeys the transmuted energy eigenproblem~\cite{Arovas1985} ${\cal H^\prime} \Psi_f = E^\prime \Psi_f$, where
\begin{subequations}
\begin{align}
\label{eq:Hp}
{\cal H^\prime} & \equiv \frac{1}{2 m} \sum_{n  = 1}^N |\pv_n - q \Av_n(\rv_n)|^2,\\
\label{eq:a-cs}
\Av_n(\rv) & \equiv \frac{\phi \hbar}{q} \sum_{n^\prime (\ne n)} \nabla_n^\prime \theta_{n,n^\prime} = \frac{\phi \hbar}{q} \!\! \sum_{n^\prime (\ne n)} \frac{\zvh \times (\rv - \rv_{n^\prime})}{|\rv - \rv_{n^\prime}|^2}.
\end{align}
\end{subequations}
${\cal H^\prime}$ has the same energy spectrum as ${\cal H}$, so for any eigenstate, including the ground state, $E^\prime = E$.

If, in Eq.~(\ref{eq:Hp}), $\Av_n(\rv_n)$ were to depend only on $\rv_n$ and not on any $\{ \rv_{n^\prime(\ne n)} \}$,
the Hamiltonian ${\cal H^\prime}$ would describe independent particles in a magnetic field $\nabla \times \Av$.
However, as shown in Eq.~(\ref{eq:a-cs}), $\Av_n$ does depend on $\{ \rv_{n^\prime(\ne n)} \}$, which implies (nonlocal) interactions between all particles. Equation (\ref{eq:a-cs}) states that $\{ \Av_n \}$ describes particles that have $\phi$ quanta of fictitious magnetic flux attached (i.e., localized at the position of every particle, each quantum carrying $2 \pi \hbar /q$ flux, where $q$ is a fictitious charge). Thus, ${\cal H}^\prime$ describes composite fermions---composed of particles obeying Fermi statistics and flux tubes~\cite{Wilczek, Jain}. This transmutation of statistics is the Hamiltonian form of Chern-Simons theory~\cite{f2}.

This formulation is effectual in incorporating the hard-core restriction. Moreover,
it opens up a pathway to a natural approximation procedure (see, e.g., Refs.~\cite{Wilczek, Jain}), which we now describe, for handling
the Chern-Simons interactions [see Eq.~(\ref{eq:a-cs})].
Instead of having an odd number of flux tubes attached to each particle, we smear the magnetic field associated with one flux tube per particle uniformly over the area of the system, and gauge-transform away the remaining flux tubes. In this so-called Average Field Approximation (AFA), the fermions are \emph{non-interacting} and subject to a \emph{homogeneous} magnetic field $ \nabla \times \Av = B \zvh$, corresponding to one quantum of magnetic flux per particle. [In the symmetric gauge, $\Av = B (y, - x, 0)/2$.]\thinspace\ In the language of directed lines, the magnetic field is equal to the number of lines per unit area $\rho_0$ times a quantum of flux (i.e., \mbox{$B = \rho_0\, 2 \pi \hbar /q$}).
In this magnetic field, the many-body ground-state has energy $E = N B \hbar q / 2 m$ and the Slater determinant wave-function takes the Vandermonde form~\cite{Laughlin1983b}
\begin{equation}
\label{eq:LLL}
\Psi^a_f(\Rv) \propto  e^{-\sum_{n = 1}^N |w_n|^2/4 \ell^2} \prod_{1 \le n < n^\prime \le N} (w_n - w_{n^\prime}),
\end{equation}
where $w_{n} \equiv x_{n} + i y_{n}$ and $\ell \equiv \sqrt{\hbar/ q B}$.
Within this approximation, the quantum ground-state energy and wave-function describe a long, noncrossing, directed line liquid.
In particular, for the free energy per unit volume of the unconstrained directed line liquid, we use the  correspondence~(\ref{eq:corr}) to obtain the result
\begin{figure}[thp]
\includegraphics[angle=0]{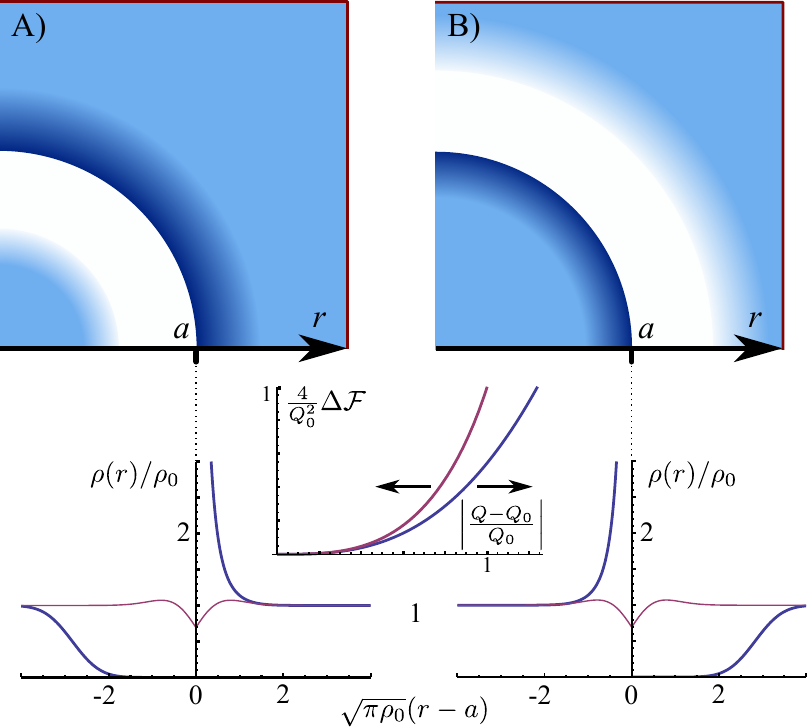}
  \caption{(\emph{Color online}) Two-dimensional one-component plasma (2DOCP) subject to a spatially extended constraint.
The electrostatic approximation predicts a gap in the mobile charge density and an accumulated surface charge for cases~(A), for which the excess mobile charge resides outside the ring, and~(B), for which the excess mobile charge is confined inside the ring.
This approximation also predicts discontinuities in the mobile-charge density, which in the exact solution are smeared out 
due to thermal fluctuations of the mobile charges around the minimum energy state. 
The exact solution leads to the density profiles $\rho(r)/\rho_0$ shown for $(Q - Q_0)/\sqrt{2 Q_0} = \pm 4$ (thick) and $0$ (thin), 
where $- Q$ and $Q_0$ ($\equiv \pi a^2 \rho_0$) are the mobile and background charges inside the ring, respectively~\cite{Jancovici1981-2,Jancovici1982-1}.
The inset shows the rescaled energy cost $(4/Q_0^2)\, \Delta {\cal F}$ [top: case (A), bottom: case (B)]~\cite{Jancovici1993}.
The curves are valid for macroscopic values of the charge deficiency, i.e., for not-too-small values of the argument, when $\Delta {\cal F}$ is dominated by electrostatic energy. 
In the text, we establish that these profiles correspond to the in-plane distribution of noncrossing directed lines due to a planar spatially extended constraint.}
  \label{fig:profile}
\end{figure}
\begin{equation}
\label{eq:g}
\frac{F_u}{L {\cal A}} = \frac{\pi}{\tau}\, T^2 \rho_0^2\,.
\end{equation}
This expression encompasses the thermodynamics of the polymer fluid. For example, for the (areal) compressibility $\kappa \equiv \left[\rho_0^{2} \,\, \partial^2  \left(F_u / L {\cal A}\right)/\partial \rho_0^2 \right]^{-1}$, we obtain $\tau/( 2 \pi T^2 \rho_0^2 )$.

As Eq.~(\ref{eq:LLL}) gives an expression for the ground-state wave-function, we proceed with our goal of calculating the
increase in the free energy $\Delta {\cal F}$ by
using the integral in Eq.~(\ref{eq:ds}). We thus make the approximation $|\Psi_b|^2 \approx |\Psi^a_f|^2$, and
compute $\Delta {\cal F} = - T \ln \int_{\cal C} d\Rv \exp\big(-U(\Rv)/T\big)$, where
\begin{equation}
- \frac{U(\Rv)}{T} \approx \ln |\Psi^a_{f}|^2 = 2 \!\!\!\!\!\!\!\!\! \sum_{\substack{1 \le n < n^\prime \le N}} 
\!\!\!\!\!\!\!\!\! \ln |\rv_n - \rv_{n^\prime}| - \pi \rho_0\!\! \sum_{n=1}^N |\rv_n|^2.
\end{equation}
Interpreted as a potential energy, $U$ describes a two-dimensional one-component plasma (2DOCP)~\cite{Laughlin1983, Jancovici1981-1}.
In the general case of the plasma, particles of (e.g., negative) charge $- e$ inhabit a uniform background that maintains overall charge neutrality, and they interact via the two-dimensional Coulomb repulsion: $e^2 \ln|\rv_n - \rv_{n^\prime}|$. 
This $U/T$ corresponds to a specific value, viz.~$2$, of the plasma coupling constant $e^2/ T$, for which the model is exactly solvable~\cite{Jancovici1981-1}.
Applied to the plasma, the measure $\int_{\cal C} d \Rv$ in $\Delta {\cal F}$ requires exactly $Q$ mobile charges to occupy the region of constraint ${\cal D}$.

This completes our reduction of the three-dimensional classical liquid of noncrossing directed lines to the classical two-dimensional one-component plasma (2DOCP);
this reduction enables us to compute physical properties such as 
the free-energy cost and the equilibrium density profiles associated with the imposition of spatially extended constraints.
The long-range planar interactions within the 2DOCP arise from the short-range interactions between the fluctuating directed lines, integrated over their length.
To calculate the free energy and the equilibrium density of the directed line liquid in the presence of a ring constraint, we analyze the 2DOCP, first
by minimizing the potential energy, and then using the exact solution of Refs.~\cite{Jancovici1981-2, Jancovici1982-1, Jancovici1993, Jancovici1981-1, Jancovici1982-2, *Jancovici1984}.
Note that the constraint mandates that there be regions of excess mobile charge and regions partially depleted of mobile charge.
To minimize the electrostatic energy, any excess mobile charge is forced up against the boundary of ${\cal D}$.
Moreover, on the other side of the boundary, a region fully depleted of mobile charge (i.e., a gap) opens up,
out to a radius within which the net charge is zero~\cite{f3}.

Similarly, at the level of electrostatic energy minimization one can readily determine the energy of the charge distribution
and, hence, the free-energy cost of the constraint $\Delta {\cal F}(Q, Q_0)$, in terms of the mobile charge $- Q$ and the background charge within ${\cal D}$, i.e., $Q_0$ ($\equiv \pi a^2 \rho_0$), both in units of $e$~\cite{Jancovici1993}; see Fig.~\ref{fig:profile}.
For the special case $Q = 0$ (i.e., ${\cal D}$ empty, corresponding to an inclusion) the free-energy cost has a simple form:
\begin{equation}
\label{eq:df}
\Delta {\cal F} = Q_0^2/4 = \pi^2 a^4 \rho_0^2 /4.
\end{equation}
This electrostatic result is significant for the noncrossing-directed-line liquid, as it differs qualitatively from the case of a noninteracting directed line liquid, for which a particle inclusion incurs a free-energy cost \emph{linear} in $a^2 \rho_0$; see~\cite{fR}. %The result in Eq.~(\ref{eq:df}) relies on the long-range nature of the effective in-plane interaction between the directed lines, in contast to particles in two dimensions interacting via a short-range potential of characteristic length $\lambda$, for which the analogous free energy cost scales as $\lambda^2 a^2 \rho_0^2$~\cite{fR}, which is greater than $a^2 \rho_0$ only if $\lambda^2 > 1/\rho_0$, i.e., the over-packed limit.

To improve upon the electrostatic approximation, we take into account the effect of thermal fluctuations on the polymer
density profile. We rely on the exact solution of the 2DOCP with the appropriate plasma coupling constant, i.e., $2$; see Refs.~\cite{Jancovici1981-2, Jancovici1982-1, Jancovici1993, Jancovici1981-1, Jancovici1982-2, *Jancovici1984}.
In the limit $a \gg 1/\sqrt{\rho_0}$, the exact density profile outside the region of constraint depends only on $a$, $\rho$ and $Q$ through
the combination $(Q - Q_0)/\sqrt{2 Q_0}$~\cite{Jancovici1981-2}.
The layer of excess mobile charge on one side forms an electrical double layer of thickness of order $1/\sqrt{\rho_0}$; see Fig.~\ref{fig:profile}c and Ref.~\cite{Jancovici1981-2}.
The region partially depleted of mobile charge does develop a soft gap, in which the charge is small but nonzero.
The mobile charge density profile progresses smoothly, according to a qualitatively error-function-like curve, through the boundary region, rapidly approaching the value that exactly compensates the background charge density $\rho_0$; see Fig.~\ref{fig:profile} and Ref.~\cite{Jancovici1981-2}.
The mobile charge density profile for the depleted side of the constraint applies to both cases, $Q > Q_0$ and $Q < Q_0$, and similarly for the excess-charge side. For $Q$ small relative to $Q_0$,
the remaining mobile charge in ${\cal D}$ forms a droplet whose shape is essentially the density profile of a system of electrons that fill the lowest Landau level: a 
flat central profile and a decay into the soft gap~\cite{Wen}. Thus, we have established that when some fixed portion of the lines of a directed line liquid is constrained to thread ${\cal D}$, the equilibrium density profile in the slice containing ${\cal D}$ is that of the correspondingly constrained 2DOCP.

Recapping our strategy, we progressed from the three-dimensional liquid of thermally fluctuating lines, to a two-dimensional quantum many-boson fluid, to a 
two-dimensional quantum many-fermion fluid coupled to a Chern-Simons gauge field, which we treated in the Average Field Approximation to obtain the filled lowest Landau-level picture.
The phenomenology of a lowest Landau level filled with noninteracting fermions is well studied, and suggests various analogous phenomena for the corresponding hard-core boson fluid.
However, as the AFA is an approximation, these analogous phenomena may be artifacts of the approximation, and we now use physical intuition to identify any such artifacts.
For example, the quantum Hall effect suggested by the AFA is one such artifact: the boson fluid does not have broken time-reversal symmetry, and therefore shows no Hall effect~\cite{Fetter1989,Dai1992, f4}.
A second artifact is suggested by the incompressibility of the filled lowest Landau level, which would incorrectly imply the incompressibility of the boson fluid~\cite{f5}.
To restore the physics of the boson fluid missed at the AFA level, the residual inter-particle interactions encoded in the corrections $\Av_n(\rv) - \Av(\rv)$ 
should be treated, e.g., via the Random Phase Approximation.
In this way, the proper compressibility of the boson fluid can be obtained, as shown in Refs.~\cite{Fetter1989,Dai1992}.
A particularly noteworthy consequence of the residual interactions $\Av_n(\rv) - \Av(\rv)$ is their ability to renormalize the effective plasma coupling constant $e^2/ T$ of the plasma analogy away from the exactly solvable case, viz., $2$. Nevertheless, we expect the general picture presented here, of the energetics and structure of the directed line liquid in the
presence of spatially extended constraints, to hold.

\noindent {\it Acknowledgments\/} -- Motivation for this work arose through discussions with Jennifer Curtis on her group's investigations of the pericellular coat using optical force probe assays, which we gratefully acknowledge. One of us (PMG) thanks for its hospitality the Aspen Center for Physics, where some of this work was undertaken. This work was supported in part by NSF DMR 09 06780 and DMR 12 07026.


\begin{thebibliography}{26}
\bibitem{f3}
For an arbitrary, simply-connected, compact region ${\cal D}$, the charge-density profile that minimizes the electrostatic energy may be addressed using a conformal mapping from this circular geometry.
\bibitem{f4}
By Hall effect we mean the occurrence of a transverse particle current in response to a longitudinal potential gradient.
\bibitem{f5}
The compressibility of the filled lowest Landau level is defined at fixed magnetic field $B$. 
In contrast, the compressibility of the boson fluid, and thus the areal compressibility of the polymer liquid $\kappa$,
is defined at fixed filling fraction $B/\rho_0 = 1$ due to the flux attachment procedure.
\end{thebibliography}
\end{document}